\documentclass{article}
\usepackage[pdftex]{graphicx}
\usepackage{graphics}
\usepackage{epsfig}
\usepackage{epstopdf}
\usepackage{hyperref}
\begin{document}

\title{Application of the Asymptotic Taylor Expansion Method to Bistable potentials}
\author{Okan \"{O}zer\thanks{Corresponding author E-mail: \textit{ozer@gantep.edu.tr}}~,~Halide Koklu and Serap Resitoglu}
\date{Department of Engineering Physics, Faculty of Engineering,
University of Gaziantep, 27310 Gaziantep, Turkey}
\maketitle

\abstract{A recent method called Asymptotic Taylor expansion
(ATEM) is applied to determine the analytical expression for
eigenfunctions and numerical results for eigenvalues of the
Schr\"{o}dinger equation for the bistable potentials. Optimal
truncation of the Taylor series gives a best possible analytical
expression for eigenfunctions and numerical result for
eigenvalues. It is shown that the results are obtained by a simple
algorithm constructed for a computer system using symbolic or
numerical calculation. It is observed that ATEM produces excellent
results appropriate with the existing literature.}

%\DeclareGraphicsExtensions{.pdf,.png,.gif,.jpg} %%  Example
%% Use this with pdfLatex:

\vskip 0.5cm

{\bf Pacs Numbers}: 03.65.-w, 03.65.Db, 03.65.Ge


\vskip 0.5cm \vskip 0.5cm \vskip 0.5cm

\

Original paper published in "Journal of Advances in Mathematical
Physics".

\

{\bf Journal-ref:}
\href{http://www.hindawi.com/journals/amp/2013/239254/} {Adv.
Math. Phys. {\bf 2013}, 239254 (2013)}

\
\newpage

%%###################################################################

\section{Introduction}\label{intro}
There is no doubt that an interesting problem in fundamental
quantum mechanics for lecturers, advanced undergraduate and
graduate students in physics and applied mathematics is to obtain
the exact solutions of the Schr\"odinger equation for any type of
potential. It is well known that the Schr\"odinger equation,
proposed by Erwin Schr\"odinger in 1926, is a second-order
differential equation that describes how the quantum state of a
physical system changes with time \cite{Schrodinger1926}. It is as
central to quantum mechanics as Newton's laws are to classical
mechanics. The solution of the Schr\"odinger equation in quantum
mechanics for a certain type potential is a fundamental component
of modern science because it leads to a function, called the state
eigenfunction of the physical system, containing all the necessary
information to describe the system and its properties. It is well
known that the quantum mechanics is established on some certain
postulates and in any introductory quantum physics textbook these
postulates can be found. One of the fundamental axiom for a bound
system is about the energy spectrum and it is required to be real
and bounded below the potential energy. The reason is that all
physical systems yield real energy values and at least one of them
is required to be the stable and the lowest-state. Another
postulate is about on the time-evolution of the quantum system,
and it is required that the probability measurement of any state
should not change with time: the probability function must not
grow up or decay in time. Thus, one needs to construct an
Hamiltonian which should be Hermitian and the time-evolution of
this Hamiltonian should also be unitary. As a result, such a
construction of an Hamiltonian provides the properties of the
quantum theory: Energy eigenvalues are found to be real and
bounded below, the norm of the eigenfunction remains constant in
time, and the symmetries of the theory are incorporated.

Unfortunately, there are not so many potentials that can be solved
exactly such as the Coulomb, harmonic oscillator, and
P\"{o}schl-Teller potentials. Since one of the source of progress
of the science depends on the study of the same problem from
different point of view, various methods have been suggested such
as numerical calculation \cite{VarshniPRA1990, NunezPRA1993}, the
variational \cite{VarshniPRA1990, StubbinsPRA1993}, the
perturbation \cite{MatthysPRA1988}, the WKB \cite{VarshniJPA1992,
Dobrovolsky2000}, the shifted 1/N expansion \cite{TangPRA1987,
RoychoudhuryJPA1998}, the Nikiforov-Uvarov (NU) \cite{Nikiforov,
GonulPS2007}, the supersymmetry (SUSY) \cite{GonulPLA2000,
QianNJP2002}, the generalized pseudospectral (GPS)
\cite{RoyPramana2005}, the asymptotic iteration method (AIM)
\cite{CiftciJPA11807} and other methods \cite{LaiPLA1980,
PatilJPA1984, RoyJPA1987, GreenePRA1976,MyhrmanJPA1983,
BechlertJPB1988} to find the approximate solutions of the
potentials that are not exactly solvable.

In this study we will apply a new formalism based on the Taylor
series expansion method, called namely Asymptotic Taylor Expansion
Method (ATEM) \cite{KocJPA2010}, to bistable potentials. These
type of potentials have been used in the quantum theory of
molecules as a crude model to describe the motion of a particle in
the presence of two centers of force \cite{RazavyAJP1980,
XieJPA2012, FelderhofPhysica2008, CiftciCPB2012, BarakatPLA2005,
SousMPLA2006, BarakatJMPLA2007}. It is mentioned in Ref.
\cite{KocJPA2010} that the Taylor Series Method
\cite{TaylorMethod, Struik1969} is an old one but it has not been
fully exploited in the analysis of both in solution of physical
and mathematical problems. It is also claimed that ATEM can also
be easily applied to solve second-order differential equations by
introducing a simple Mathematica \cite{Mathematica} computer
program. Therefore, we focus on the solution of the eigenvalue
problems of some type of bistable potentials by using ATEM, in
this paper.

The organization of the paper is as follows: In Section
\ref{ForATEM}, we present a brief outline of the method ATEM. In
Section \ref{applications}, the eigenvalues of the bistable
potentials are determined by using ATEM. Section \ref{conclusion}
is devoted to a conclusion.

\section{Formalism of Asymptotic Taylor Expansion Method}\label{ForATEM}
In this section, we present the solution of the Schr\"odinger-type
equations by modifying Taylor series expansion with the aid of a
finite sequence instead of an infinite sequence and its
termination possessing the property of quantum mechanical wave
function. It is well known in quantum mechanics, the bound state
energy of an atom is quantized and eigenvalues are discrete.
Additionally, for each eigenvalues there exist one or more
eigenfunctions. If one considers the solution of the
Schr\"{o}dinger equation, discrete eigenvalues of the problem are
mainly under investigation. The first main result of this
conclusion gives necessary and sufficient conditions for the
termination of the Taylor series expansion of the wave function.

Following the notation in \cite{KocJPA2010}, one can consider the
Taylor series expansion of a function $f(x)$ about the point $a$:
\begin{eqnarray}\label{a1}
f(x) &=&f(a)+(x-a)f^{\prime }(a)+\frac{1}{2}(x-a)^{2}f^{\prime
\prime }(a)+\frac{1}{6}(x-a)^{3}f^{(3)}(a)+\cdots \nonumber
\\
&=&\sum_{n=0}^{\infty}\frac{(x-a)^{n}}{n!}f^{(n)}(a)
\end{eqnarray}
where $f^{(n)}(a)$ is the $n^{th}$ derivative of the function at
$a$. Taylor series specifies the value of a function at one point,
$x$, in terms of the value of the function and its derivatives at
a reference point $a$. Expansion of the function $f(x)$ about the
origin ($a=0$), is known as Maclaurin's series and it is given by,
\begin{eqnarray}\label{a11}
f(x) &=&f(0)+xf^{\prime }(0)+\frac{1}{2}x^{2}f^{\prime \prime
}(0)+\frac{1}{6}x^{3}f^{(3)}(0)+\cdots \nonumber
\\
&=&\sum_{n=0}^{\infty}\frac{x^{n}}{n!}f^{(n)}(0).
\end{eqnarray}
Here one can develop a method to solve a second order linear
differential equation of the form:
\begin{equation}\label{a2}
f^{\prime \prime }(x)=p_{0}(x)f^{\prime }(x)+q_{0}(x)f(x).
\end{equation}
It is seen that the higher order derivatives of the $f(x)$ can be
obtained in terms of the $f(x)$ and $f^{\prime }(x)$ by
differentiating (\ref{a2}). Then, higher order derivatives of
$f(x)$ are given by
\begin{equation}\label{a3}
f^{(n+2)}(x)=p_{n}(x)f^{\prime }(x)+q_{n}(x)f(x)
\end{equation}
where
\begin{eqnarray}\label{a4}
p_{n}(x) &=&p_{0}(x)p_{n-1}(x)+p_{n-1}^{\prime }(x)+q_{n-1}(x),
\nonumber
\\
q_{n}(x) &=&q_{0}(x)p_{n-1}(x)+q_{n-1}^{\prime }(x).
\end{eqnarray}
At this point, one can observe that the eigenvalues and
eigenfunctions of the Schr\"{o}dinger-type equations can
efficiently be determined by using ATEM. To this end, the
recurrence relations (\ref{a4}) allows one to determine algebraic
exact or approximate analytical solution of (\ref{a2}) under some
certain conditions. Let us now substitute (\ref{a4}) into
(\ref{a1}) to obtain the function that is related to the wave
function of the corresponding Hamiltonian:
\begin{equation}\label{a5}
f(x)=f(0)\left( 1+\sum_{n=2}^{m}q_{n-2}(0)\frac{x^{n}}{n!}\right)
+
f^{\prime}(0)\left(1+\sum_{n=2}^{m}p_{n-2}(0)\frac{x^{n}}{n!}\right).
\end{equation}
After all, one can now obtain useful formalism of the Taylor
expansion method. This form of the Taylor series can also be used
to obtain series solution of the second order differential
equations. In the solution of the eigenvalue problems, truncation
of the the asymptotic expansion to a finite number of terms is
useful. If the series optimally truncated at the smallest term
then the asymptotic expansion of series is known as
superasymptotic \cite{Boyd1999}, and it leads to the determination
of eigenvalues with minimum error.

Since the improper sets of boundary conditions may produce
nonphysical results, arrangement of the boundary conditions for
different problems becomes very important: When only odd or even
power of $x$ collected as coefficients of $f(0)$ or
$f^{\prime}(0)$ and vice versa, the series is truncated at $n=m$
then an immediate practical consequence of these conditions is
obtained for $q_{m-2}(0)=0$ or $p_{m-2}(0)=0$. In this way, the
series truncates at $n=m$ and one of the parameter in the
$q_{m-2}(0)$ or $p_{m-2}(0)$ belongs to the spectrum of the
Schr\"{o}dinger equation. Therefore eigenfunction of the equation
becomes a polynomial of degree $m$. Otherwise the spectrum of the
system can be obtained as follows: In a quantum mechanical system
eigenfunction of the system is discrete. Therefore in order to
terminate the eigenfunction $f(x)$ we can concisely write that
\begin{eqnarray}\label{a5y}
q_{m}(0)f(0)+p_{m}(0)f^{\prime }(0) &=&0 \nonumber
\\
q_{m-1}(0)f(0)+p_{m-1}(0)f^{\prime }(0) &=&0
\end{eqnarray}
eliminating $f(0)$ and $f^{\prime }(0)$ we obtain
\begin{equation}\label{a5z}
q_{m}(0)p_{m-1}(0)-p_{m}(0)q_{m-1}(0)=0
\end{equation}
again one of the parameter in the equation related to the
eigenvalues of the problem.

It has been stated that the ATEM reproduces exact solutions to
many exactly solvable differential equations and these equations
can be related to the Schr\"{o}dinger equation. It is observed
that the process presented in ATEM is iterative and the number of
iteration is given by $m$. The method can be applied to the
Schr\"{o}dinger equation with any type of potential as following:
Using a computer program, one first sets up the iteration number
$m$, say $m=30$, to obtain the result. Then, setting $m=40$,
another result is obtained. And this procedure is repeated for
different $m$ values leading to different results. Finally, one
can compare the results for each case till desired digits. If the
values of the eigenvalue reach their asymptotic values, then one
can choose the corresponding $m$ value, and truncate the iteration
for next calculations. For instance, if one can obtain the values
of the eigenvalues for $m=60$, first few of them -first eight
eigenvalues for example- will reach automatically to their
asymptotic values. The following comment on the function is
considerable: For such a solution it may be suitable to take sum
of the first eight terms in (6).

\section{Applications}\label{applications}
We shall illustrate here that equation (\ref{a2}) with conditions
(\ref{a5y}) and (\ref{a5z}) gives a complete solution for some
important Schr\"{o}dinger-type problems. Through a concrete
example we explore the solution of Schr\"{o}dinger equation
($\hbar^{2}=2m=1$)
\begin{equation}\label{SchrodingerEq}
\left( -\frac{d^2}{dx^2}+V(x) \right)\psi(x)=E\psi(x)
\end{equation}
for the harmonic oscillator potential in one dimension
$V(x)=x^{2}$ given as
\begin{equation}\label{SHOHam}
\left( -\frac{d^2}{dx^2}+x^{2} \right)\psi(x)=E\psi(x)
\end{equation}
\begin{figure}[t]
\begin{center}
\includegraphics[width=0.6\textwidth]{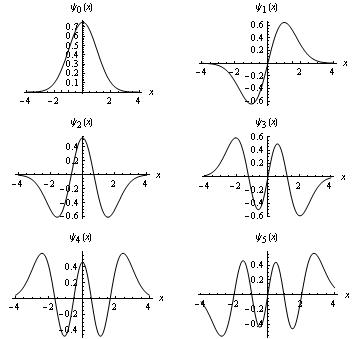}
\caption {The first six states normalized eigenfunctions of the
Harmonic potential given in Eq. (\ref{SHOHam}).}
\label{FigureSHOWFs}
\end{center}
\end{figure}
In the limit of large $x$, the asymptotic solutions of
(\ref{SHOHam}) can be taken as any power of $x$ times a decreasing
Gaussian function to satisfy the quantum mechanics postulates.
With this in mind, one can write the 'unnormalized' wavefunctions
as
\begin{equation}\label{SHOpsi0}
\psi(x)=exp \left( -\frac{x^{2}}{2} \right) f(x)
\end{equation}
where the functions $f(x)$ are to be found by means of the
iteration procedure given above. Substituting (\ref{SHOpsi0}) into
(\ref{SHOHam}), one obtains
\begin{equation}\label{SHOLag}
L(x)=-f''(x)+2 x f'(x) + (1-E) f(x)=0.
\end{equation}
Comparing Eqns. (\ref{a2}) and (\ref{SHOLag}) one can deduce that
\begin{equation}\label{SHOp0q0}
p_{0}(x)=2x~~,~~ q_{0}(x)=1-E
\end{equation}
Following the procedure given in Eq. (\ref{a5z}) yields the exact
eigenvalues of the Harmonic oscillator potential:
\begin{equation}\label{SHOEnergy}
E_{n}=2n+1~~,~~n=0,1,2,3\ldots
\end{equation}
and the normalized eigenfunctions of the Schr\"{o}dinger equation
in Eq. (\ref{SHOHam}) are shown in Figure \ref{FigureSHOWFs}.
%for $m=10$, then one obtains the first nine energy states as shown
%in Table \ref{SHOTable}.
%\begin{table}
%\caption{The first nine eigen states of the potential in Eq.
%(\ref{SHOHam}). \label{SHOTable}}
%\begin{tabular}{cccccccccc}
%\hline
%$E_{0}$ & $E_{1}$ & $E_{2}$ & $E_{3}$ & $E_{4}$ & $E_{5}$ & $E_{6}$ & $E_{7}$ & $E_{8}$ & $E_{9}$ \\
%\hline \hline
%$1$  &   $3$   &   $5$   & $7$  & $9$ & $11$ & $13$ & $15$ & $17$ & $19$ \\
%\hline
%\end{tabular}
%\end{table}

As a second application, we consider the quartic anharmonic
oscillator potential that has been a great deal of interest in the
analytical and numerical investigation of the one-dimensional
anharmonic oscillators because of their importance in molecular
vibrations \cite{HsuePRD1984} as well as in solid state physics
\cite{FlessasJPA1983, BonhamJCP1996} and quantum field theories
\cite{Reed1978}. Schr\"{o}dinger equation (\ref{SchrodingerEq})
for the one-dimensional quartic anharmonic potential $V(x)=x^{2}+g
x^{4}$ is written as
\begin{equation}\label{HamQuarticPot}
\left( -\frac{d^2}{dx^2}+x^{2}+g x^{4} \right)\psi(x)=E\psi(x).
\end{equation}
where $g>0$. The solution of Eq. (\ref{HamQuarticPot}) has been
always studied to test accuracy and efficiency of the different
methods proposed in \cite{CiftciJPA11807, AquinoJMC1995,
BacusJPA1995, GonulMPLA2005, KocArXiv2010}.
%We should note here that the study in Ref. \cite{KocArXiv2010} applies the ATEM to
%find the eigenvalues and eigenfunctions of the quadratic
%anharmonic potential only for $g=0.1$.
Now, we introduce the asymptotic solutions of Eq.
(\ref{HamQuarticPot}) as
\begin{equation}\label{QuarticPsi}
\psi(x)=e^{-\alpha\frac{x^2}{2}-\beta\frac{x^4}{4}}f(x)
\end{equation}
and Eq. (\ref{HamQuarticPot}) can now be written as
%\begin{equation}\label{QuarticLag}
%L(x)=-f''(x)+2x (\alpha+\beta x^{2}) f'(x) +
%\left(\alpha-E+(1-\alpha^{2}+3\beta) x^{2} +
%(g-2\alpha\beta)x^{4}-\beta x^{6}\right)f(x)=0.
%\end{equation}
\begin{eqnarray}\label{QuarticLag}
L(x)=&-&f''(x)+2 (\alpha x+\beta x^{3}) f'(x) \nonumber
\\
&+& \left(\alpha-E+(1-\alpha^{2}+3\beta) x^{2} +
(g-2\alpha\beta)x^{4}-\beta x^{6}\right)f(x)=0. \nonumber
\\
\end{eqnarray}
Comparing Eqns. (\ref{a2}) and (\ref{QuarticLag}) one can deduce
that
\begin{equation}\label{Quarticp0q0}
p_{0}(x)=2(\alpha x+\beta x^{3})~~,~~
q_{0}(x)=\alpha-E+(1-\alpha^{2}+3\beta) x^{2} +
(g-2\alpha\beta)x^{4}-\beta x^{6}.
\end{equation}
By the aid of computer program, one can calculate the eigenvalues
$E_n$ and the corresponding eigenfunctions $f(x)$ for a range of
$g$ value, changing from $0.01$ to $100$, using number of
iterations $m = 20, 20, 40, 60, 80, 100, 120$. The term
"asymptotic" means the function approaching to a given value as
the iteration number $m$ tends to infinity.

\begin{table}
\caption{Eigenvalues of the potential in Eq. (\ref{HamQuarticPot})
for a range of $g$ value. For ATEM results, number of iterations
for each $g$ value is set to $120$. For SUSY results, numbers in
parenthesis ($N$) denotes the perturbation order used by authors
in Ref. \cite{GonulMPLA2005}.}
\begin{center}\label{QT1}
\begin{tabular}{c|ccc}
\hline $g$ & $E_{ATEM}$ & $E_{SUSY} \cite{GonulMPLA2005}$
& $E_{exact} \cite{BanerjeePRC1978}$ \\
\hline \hline $0.01$ & $
\begin{array}{c}
1.007373 \\
3.036525 \\
5.093939 \\
7.178573
\end{array}
$ & $
\begin{array}{c}
1.00737 (N=4) \\
3.03653 (N=8) \\
5.09609 (N=15) \\
7.19832 (N=15)
\end{array}
$ & $
\begin{array}{c}
1.007373 \\
3.036525 \\
5.093939 \\
7.178573
\end{array}
$ \\
\hline $0.05$ & $
\begin{array}{c}
1.034729 \\
3.167225 \\
5.417261 \\
7.770271
\end{array}
$ & $
\begin{array}{c}
1.03473 (N=4) \\
3.16723 (N=8) \\
5.42404 (N=15) \\
7.83995 (N=15)
\end{array}
$ & $
\begin{array}{c}
1.034729 \\
3.167225 \\
5.417261 \\
7.770271
\end{array}
$ \\
\hline $0.1$ & $
\begin{array}{c}
1.065286 \\
3.306872 \\
5.747959 \\
8.352678
\end{array}
$ & $
\begin{array}{c}
1.06528 (N=4) \\
3.30687 (N=8) \\
5.75694 (N=15) \\
8.45913 (N=15)
\end{array}
$ & $
\begin{array}{c}
1.065286 \\
3.306872 \\
5.747959 \\
8.352678
\end{array}
$ \\
\hline $0.5$ & $
\begin{array}{c}
1.2418541 \\
4.051932 \\
7.396900 \\
11.11515
\end{array}
$ & $
\begin{array}{c}
1.24118 (N=4) \\
4.05171 (N=8) \\
7.40489 (N=15) \\
11.3415 (N=15)
\end{array}
$ & $
\begin{array}{c}
1.2418541 \\
4.051932 \\
7.396900 \\
11.11515
\end{array}
$ \\
\hline $1$ & $
\begin{array}{c}
1.392352 \\
4.648813 \\
8.655049 \\
13.15680
\end{array}
$ & $
\begin{array}{c}
1.39017 (N=4) \\
4.64784 (N=8) \\
8.65908 (N=15) \\
13.4524 (N=15)
\end{array}
$ & $
\begin{array}{c}
1.392352 \\
4.648813 \\
8.655049 \\
13.15680
\end{array}
$ \\
\hline $10$ & $
\begin{array}{c}
2.449174 \\
8.599003 \\
16.63592 \\
25.80627
\end{array}
$ & $
\begin{array}{c}
2.42910 (N=4) \\
8.58582 (N=8) \\
16.6188 (N=15) \\
26.4698 (N=15)
\end{array}
$ & $
\begin{array}{c}
2.449174 \\
8.599003 \\
16.63592 \\
25.80627
\end{array}
$ \\
\hline $100$ & $
\begin{array}{c}
4.999410 \\
17.83000 \\
34.87117 \\
54.36576
\end{array}
$ & $
\begin{array}{c}
4.93770 (N=4) \\
17.7864 (N=8) \\
34.8238 (N=15) \\
55.4001 (N=15)
\end{array}
$ & $
\begin{array}{c}
4.999418 \\
17.83019 \\
34.87398 \\
54.38529
\end{array}
$ \\
\hline
\end{tabular}
\end{center}
\end{table}

We present our results carried out for a range of $g$ values in
Table \ref{QT1} with $7$ significant digits and they are compared
with those of supersymmetric perturbation approach by Ref.
\cite{GonulMPLA2005} and the ones computed numerically by Ref.
\cite{BanerjeePRC1978}. In our calculations, we set $m=120$,
$\alpha=4$ and $\beta=0$. It is observed that there is remarkable
agreement in the whole range of values for all quantum states for
different $g$ values with results of \cite{BanerjeePRC1978} except
$g=100$.

\begin{figure}
\begin{center}
\includegraphics[width=0.6\textwidth]{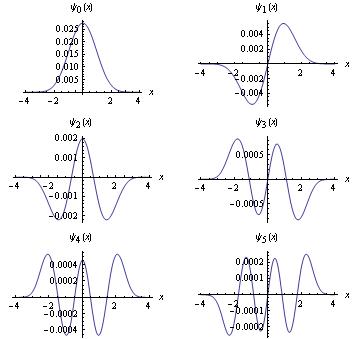}
\caption {The first six states normalized eigenfunctions of the
quartic potential given in Eq. (\ref{HamQuarticPot}) for $g=0.1$.}
\label{FigureQuarticWFs}
\end{center}
\end{figure}

We also present and compare our results for $g=0.1$ with those of
Bacus \textit{et al.} \cite{BacusJPA1995} in Table \ref{QT2} with
20 significant digits.
\begin{table}
\caption{The first ten eigenvalues of the potential in Eq.
(\ref{HamQuarticPot}) for $g=0.1$. No of iteration is set to
$m=120$.}
\begin{center}\label{QT2}
\begin{tabular}{ccc}
\hline
$n$ & $E_{ATEM}$ & $E \cite{BacusJPA1995}$ \\
\hline \hline $
\begin{array}{c}
0\\
1\\
2\\
3\\
4\\
5\\
6\\
7\\
8\\
9
\end{array}
$ & $
\begin{array}{c}
1.065~285~509~543~717~701\\
3.306~ 872~ 013~ 152~ 913~ 680\\
5.747~ 959~ 268~ 833~ 563~ 228\\
8.352~ 677~ 825~ 785~ 754~ 350\\
11.098~ 595~ 622~ 633~ 043~ 333\\
13.969~ 926~ 197~ 742~ 799~ 089\\
16.954~ 794~ 686~ 144~ 150~ 972\\
20.043~ 863~ 604~ 188~ 462~ 801\\
23.229~ 552~ 179~ 939~ 290~ 112\\
26.505~ 554~ 752~ 536~ 617~ 968
\end{array}
$ & $
\begin{array}{c}
1.065~285~ 509~ 543~ 717~ 688\\
3.306~ 872~ 013~ 152~ 913~ 507\\
5.147~ 959~ 268~ 833~ 563~ 304\\
8.352~ 677~ 825~ 785~ 754~ 712\\
11.098~ 595~ 622~ 633~ 043~ 011\\
13.969~ 926~ 197~ 742~ 799~ 300\\
16.954~ 794~ 686~ 144~ 151~ 337\\
20.043~ 863~ 604~ 188~ 461~ 233\\
23.229~ 552~ 179~ 939~ 289~ 070\\
26.505~ 554~ 752~ 536~ 617~ 417
\end{array}
$ \\
\hline
\end{tabular}
\end{center}
\end{table}
%\begin{table}
%\caption{The first ten eigenvalues of the potential in Eq.
%(\ref{HamQuarticPot}) for $g=0.1$. No of iteration is set to
%$m=120$.}
%\begin{center}\label{QT2}
%\begin{tabular}{ccc}
%\hline
%$n$ & $E_{ATEM}$ & $E \cite{BacusJPA1995}$ \\
%\hline \hline $
%\begin{array}{c}
%0\\
%1\\
%2\\
%3\\
%4\\
%5\\
%6\\
%7\\
%8\\
%9
%\end{array}
%$ & $
%\begin{array}{c}
%1.065~285~509~543~717~701~319~565~094~31\\
%3.306~ 872~ 013~ 152~ 913~ 680~ 755~ 773~ 384~ 59\\
%5.747~ 959~ 268~ 833~ 563~ 228~ 895~ 400~ 243~ 35\\
%8.352~ 677~ 825~ 785~ 754~ 350~ 818~ 024~ 249~ 75\\
%11.098~ 595~ 622~ 633~ 043~ 333~ 848~ 945~ 621~ 9\\
%13.969~ 926~ 197~ 742~ 799~ 089~ 951~ 404~ 312~ 2\\
%16.954~ 794~ 686~ 144~ 150~ 972~ 495~ 481~ 255~ 5\\
%20.043~ 863~ 604~ 188~ 462~ 801~ 403~ 147~ 750~ 5\\
%23.229~ 552~ 179~ 939~ 290~ 112~ 997~ 696~ 269~ 3\\
%26.505~ 554~ 752~ 536~ 617~ 968~ 416~ 976~ 509~ 6
%\end{array}
%$ & $
%\begin{array}{c}
%1.065~285~ 509~ 543~ 717~ 688~ 857~ 091~ 628~ 79\\
%3.306~ 872~ 013~ 152~ 913~ 507~ 128~ 121~ 684~ 69\\
%5.147~ 959~ 268~ 833~ 563~ 304~ 733~ 503~ 118~ 48\\
%8.352~ 677~ 825~ 785~ 754~ 712~ 155~ 257~ 734~ 64\\
%11.098~ 595~ 622~ 633~ 043~ 011~ 086~ 458~ 749~ 3\\
%13.969~ 926~ 197~ 742~ 799~ 300~ 973~ 433~ 956~ 8\\
%16.954~ 794~ 686~ 144~ 151~ 337~ 692~ 616~ 508~ 8\\
%20.043~ 863~ 604~ 188~ 461~ 233~ 641~ 421~ 107~ 4\\
%23.229~ 552~ 179~ 939~ 289~ 070~ 647~ 087~ 434~ 3\\
%26.505~ 554~ 752~ 536~ 617~ 417~ 469~ 503~ 006~ 7
%\end{array}
%$ \\
%\hline
%\end{tabular}
%\end{center}
%\end{table}
The function $f(x)$, for $g=0.1$, $n = 5$ state for different
values of $m$ is found as followings:
\begin{eqnarray}\label{FxFunctionsQuartic}
m=20; f(x) = x &-&1.3425 x^{3}-1.28331 x^{5}-0.249794
x^{7}+0.126907 x^{9}
\nonumber \\
&+&0.0974402 x^{11}+0.0325753 x^{13}+0.00674249 x^{15}+0.000816039
x^{17}
\nonumber \\
m=40; f(x) = x&-&0.35941 x^{3}-0.998775 x^{5}-0.578949
x^{7}-0.161047 x^{9}
\nonumber \\
&-&0.0125032 x^{11}+0.0088751 x^{13}+0.00478244 x^{15}+0.00143729
x^{17}
\nonumber\\
m=80; f(x) = x&-&0.328321 x^3-0.980318 x^5-0.581291 x^{7}-0.167846
x^9
\nonumber \\
&-&0.0162743 x^{11}+0.00762613 x^{13}+0.00449722 x^{15}+0.00139268
x^{17}
\nonumber\\
m=120; f(x) = x&-&0.328321 x^{3}-0.980318 x^{5}-0.581291
x^{7}-0.167846 x^9
\nonumber \\
&-&0.0162743 x^{11}+0.00762613 x^{13}+0.00449722 x^{15}+0.00139268
x^{17}\nonumber
\\
\end{eqnarray}
For the first six states, the plot of the normalized wave
functions for $g=0.1$ are given in Figure \ref{FigureQuarticWFs}.
As a last example, we consider the symmetric bistable potential
defined in Ref.\cite{BorgesPA2010} as
\begin{equation}\label{VsPot}
V(x)=x^{6}-2 x^{4}-2 x^{2}+1~~.
\end{equation}
Inserting the potential function into Schr\"{o}dinger equation in
Eq. (\ref{SchrodingerEq}) and using the \textit{ansatz} wave
function of the form defined in Eq. (\ref{QuarticPsi}) one obtains
\begin{eqnarray}\label{VsLag}
L(x)=&-&f''(x) + 2 \left( \alpha x+\beta  x^{3} \right) f'(x)
\nonumber
\\
&+&\left(1+\alpha-2E-(2+\alpha^{2}-3\beta)x^{2}-2(1+\alpha
\beta)x^{4}+(1-\beta^{2})x^{6} \right) f(x)=0.\nonumber
\\
\end{eqnarray}
Following the procedure given in Eq. (\ref{a5z}) for $m=120$,
$\alpha=4$ and $\beta=1$ yields the eigenvalues of the symmetric
bistable potential in Eq. (\ref{VsPot}). Our results are presented
and compared with the values found by the variational
supersymmetric method ($E_{VSQM}$)  \cite{BorgesPA2010},
state-dependent diagonalization method ($E_{SDD}$)
\cite{SoPhysicaA2000} and by direct numerical integration
($E_{Exact}$) \cite{BorgesPA2010}, in Table \ref{VsTable}. The
ATEM results are in a very good agreement, by low percent errors,
for the all values of energies with the ones obtained by numerical
calculation. For lower percent errors, the iteration number $m$
must be increased.
\begin{table}
\caption{Comparison of energy eigenvalues of the potential in Eq.
(\ref{VsPot}). For ATEM results, number of iterations is set to
$m=120$.}
\begin{center}\label{VsTable}
\begin{tabular}{cccc}
\hline $E_{ATEM}$ & $E_{VSQM} \cite{BorgesPA2010}$ & $E_{SDD}
\cite{SoPhysicaA2000}$
& $E_{Exact} \cite{BorgesPA2010}$ \\
\hline \hline $
\begin{array}{c}
0 \\
0.4229446 \\
2.314913 \\
4.503779 \\
7.175475 \\
10.27788 \\
13.75855 \\
17.58421 \\
21.72951 \\
26.17305
\end{array}
$ & $
\begin{array}{c}
0 \\
0.4238512 \\
2.319117 \\
4.571588 \\
7.101165 \\
9.861245 \\
12.82074 \\
15.95720 \\
19.25351 \\
22.69614
\end{array}
$ & $
\begin{array}{c}
0 \\
0.4229446 \\
2.314913 \\
4.503779 \\
7.175475 \\
10.27789 \\
13.75855 \\
17.58420 \\
21.72942 \\
26.17370
\end{array}
$ & $
\begin{array}{c}
0 \\
0.4229511 \\
2.314925 \\
4.503822 \\
7.175509 \\
10.27797 \\
13.75861 \\
17.58434 \\
21.72951 \\
26.17391
\end{array}
$ \\
\hline
\end{tabular}
\end{center}
\end{table}

\section{Conclusion}\label{conclusion}
An approximate method based on the asymptotic Taylor series
expansion of a function and its fundamental features are
presented. It is observed that the method is applicable for
obtaining both eigenvalues and eigenfunctions of the
Schr\"{o}dinger-type equations. After applying the method to the
one-dimensional Harmonic oscillator potential, it is shown that
the approach gives accurate results for eigenvalue problems of
some certain type bistable potentials. It is thought that the
approach opens the way to the treatment of the Schrodinger
equation including large class of potentials of practical
interest. As a future study, the method can be developed and
applied to the non Hermitian systems and QES potentials. It would
also be interesting the direct application of the method to the
Fokker–Planck equation for quasi exactly solvable bistable
potentials when the drift coefficient has a non-polynomial nature.
Studies along this line are in progress.

\section*{Acknowledgements}
%the Scientific and Technical Research Council of Turkey (T\"{U}B\.{I}TAK).
This work is supported by the Research Fund Unit of Gaziantep
University. O. \"{O}zer would like to thank Dr G. L\'evai (ATOMKI,
Debrecen-Hungary) for supporting some papers used in this study
and it is a pleasure to thank Dr R. Koc for valuable conversations
and comments.
%H. Ahmadov
%The author is also very much indebted to Dr X Z
%the referee
%for useful comments and suggestions.
%comments.
%We are thankful to the anonymous referees for several suggestions
%which led to fine tuning some of the results presented in this paper


\begin{thebibliography}{99}
%%%%%%%%%%%%%%%%%%%%%%%%%%%%%%%%%%%%%%%%%%%%%%%%%%%%%%%%%%%%%
% Some macros are available for the bibliography:
%   o for general use
%      \JL : general journals          \andvol : Vol (Year) Page
%   o for individual journal
%    \AJ   : Astrophys. J.           \NC         : Nuovo Cim.
%    \ANN  : Ann. of Phys.           \NPA, \NPB  : Nucl. Phys. [A,B]
%    \CMP  : Commun. Math. Phys.     \PLA, \PLB  : Phys. Lett. [A,B]
%    \IJMP : Int. J. Mod. Phys.      \PRA - \PRE : Phys. Rev. [A-E]
%    \JHEP : J. High Energy Phys.    \PRL        : Phys. Rev. Lett.
%    \JMP  : J. Math. Phys.          \PRP        : Phys. Rep.
%    \JP   : J. of Phys.             \PTP        : Prog. Theor. Phys.
%    \JPSJ : J. Phys. Soc. Jpn.      \PTPS       : Prog. Theor. Phys. Suppl
% Usage:
%   \PR{D45,1990,345}          ==> Phys.~Rev.\ \textbf{D45} (1990), 345
%   \JL{Nature,418,2002,123}   ==> Nature \textbf{418} (2002), 123
%   \andvol{B123,1995,1020}    ==> \textbf{B123} (1995), 1020
%%%%%%%%%%%%%%%%%%%%%%%%%%%%%%%%%%%%%%%%%%%%%%%%%%%%%%%%%%%%%

\bibitem{Schrodinger1926} Schr\"odinger E 1926 Ann. Phys. \textbf{79} 361

\bibitem{VarshniPRA1990} Varshni Y P 1990 Phys. Rev. A \textbf{41} 4682

\bibitem{NunezPRA1993} Nunez M A 1993 Phys. Rev. A \textbf{47} 3620

\bibitem{StubbinsPRA1993} Stubbins C 1993 Phys. Rev. A \textbf{48} 220

\bibitem{MatthysPRA1988} Matthys P and Meyer H D 1988 Phys. Rev. A \textbf{38} 1168

\bibitem{VarshniJPA1992} Varshni Y P 1992 J. Phys. A \textbf{25}, 5761

\bibitem{Dobrovolsky2000}Dobrovolsky G A and Tutik R S 2000 J. Phys. A: Math. Gen.
\textbf{33} 6593

\bibitem{TangPRA1987} Tang A Z and Chan F T 1987 Phys. Rev. A \textbf{35} 911

\bibitem{RoychoudhuryJPA1998} Roychoudhury R K and Varshni Y P 1998 J. Phys. A \textbf{21} 3025

\bibitem{Nikiforov} Nikiforov A F and Uvarov V B 1988 \textit{Special Functions of
Mathematical Physics} (Basel: Birkhauser)

\bibitem{GonulPS2007}Gonul Bu and Koksal K 2007 Phys. Scr. \textbf{75} 686

\bibitem{GonulPLA2000} Gonul Bu, Ozer O, Cancelik Y and Kocak M 2000 Phys. Lett. A
\textbf{275} 238

\bibitem{QianNJP2002}Qian S W, Huang B W and Gu Z Y 2002 New J. Phys. \textbf{4} 13

\bibitem{RoyPramana2005} Roy A K 2005 Pramana J. Phys. \textbf{65} 1

\bibitem{CiftciJPA11807} Ciftci H, Hall R L and Saad N 2003 J. Phys. A: Math. Gen. \textbf{36} 11807

\bibitem{LaiPLA1980} Lai C S and Lin W C 1980 Phys. Lett. A \textbf{78} 335;

\bibitem{PatilJPA1984} Patil S. H. 1984 J. Phys. A: Math. Gen. \textbf{17} 575;

\bibitem{RoyJPA1987} Roy B and Roychoudhury R 1987 J. Phys. A: Math. Gen. \textbf{20}
3051;

\bibitem{GreenePRA1976} Greene R L and Aldrich C 1976 Phys. Rev. A \textbf{14} 2363;

\bibitem{MyhrmanJPA1983} Myhrman U 1983 J. Phys. A: Math. Gen. \textbf{16} 263;

\bibitem{BechlertJPB1988} Bechlert A and Bühring W 1988 J. Phys. B: At. Mol. Opt. Phys.
\textbf{21} 817

\bibitem{KocJPA2010} Koc R and Sayin S 2012 J. Phys. A: Math. Gen. \textbf{43}
455203

\bibitem{RazavyAJP1980} Razavy M 1980 Am. J. Phys. \textbf{48} 285

\bibitem{XieJPA2012} Xie Q-T 2012 J. Phys. A: Math. Theor. \textbf{45} 175302

\bibitem{FelderhofPhysica2008} Felderhof B U 2008 Physica A \textbf{387} 5017

\bibitem{CiftciCPB2012} Ciftci H, Ozer O and Roy P 2012 Chin. Phys. B
\textbf{21} 010303

\bibitem{BarakatPLA2005} Barakat T 2005 Phys. Lett. A \textbf{344} 411

\bibitem{SousMPLA2006} Sous A J 2006 Mod. Phys. Lett. A \textbf{21} 1675

\bibitem{BarakatJMPLA2007} Barakat T and Al-Dossary O M 2007 Int. J. Mod. Phys. A \textbf{22}
2003

\bibitem{TaylorMethod} Taylor Brook 1715 \textit{Methodus Incrementorum Directa et Inversa}
(Direct and Reverse Methods of Incrementation) (London)

\bibitem{Struik1969} Struik D J 1969 \textit{A Source Book in Mathematics}
(Translated into English) (Cambridge, Massachusetts: Harvard
University Press)

\bibitem{Mathematica} Wolfram Research, Inc., Mathematica, Version 8.0, Champaign, IL (2010).

\bibitem{Boyd1999} Boyd J P 1999 Acta Appl. Math. \textbf{56} 1

\bibitem{HsuePRD1984} Hsue C and Chern J L 1984 Phys. Rev. D \textbf{29} 643

\bibitem{FlessasJPA1983} Flessas G P, Whitehead R R and Rigas A 1983 J. Phys. A: Math. Gen. \textbf{16} 85

\bibitem{BonhamJCP1996} Bonham R A and Su L S 1996 J. Chem. Phys.\textbf{45} 2827

\bibitem{Reed1978} Reed M and Simon B 1978 \textit{Methods of Modern Mathematical Physics, IV
Analysis of Operators} (New York, Academic)

\bibitem{AquinoJMC1995} Aquino N 1995 J. Math. Chem. \textbf{18} 349

\bibitem{BacusJPA1995} Bacus B, Meurice Y and Soemadi A 1995 J. Phys. A: Math. Gen. \textbf{28} L381

\bibitem{GonulMPLA2005} Gonul Bu, Celik N and Olgar E 2005  Mod. Phys. Lett. A. \textbf{20} 1683

\bibitem{KocArXiv2010} Koc R and Olgar E 2010 Preprint math-ph/1008.0697

\bibitem{BanerjeePRC1978} Banerjee K 1978 Proc. R. Soc. A \textbf{364} 265

\bibitem{BorgesPA2010} Borges G R P, Filho E D and Ricotta R M 2010 Physica A \textbf{389} 3892

\bibitem{SoPhysicaA2000} So F and Liu K L 2000 Physica A \textbf{277} 335
\end{thebibliography}
\end{document}